\newcommand{\equi}{\mbox{$\leftrightarrow$}}
\newcommand{\iffi}{\mbox{$\quad\Leftrightarrow\quad$}}
\newcommand{\imp}{\to}
\newcommand{\Imp}{\mbox{$\quad\Rightarrow\quad$}}
\newcommand{\nvdash}{\mbox{$\;\not\vdash\;$}}
\newcommand{\rwedge}{\mbox{\raisebox{.4ex}{$\bigwedge$}}}
\newtheorem{theorem}{Theorem}
\newtheorem{coroll}[theorem]{Corollary}
\newtheorem{definition}[theorem]{Definition}
\newtheorem{fact}[theorem]{Fact}
\newtheorem{lemma}[theorem]{Lemma}

\documentclass{tlp}
\title[Intermediate Logics and Strongly Equivalent Programs]
{Characterization of Strongly Equivalent Logic Programs in
Intermediate Logics}
\author[D.H.J. de Jongh and A. Hendriks]
{DICK DE JONGH and LEX HENDRIKS\\
Institute of Logic, Language and Computation, University of Amsterdam\\
Plantage Muidergracht 24\\
1018 TV Amsterdam\\
The Netherlands\\
\email{dickdj@science.uva.nl}, \email{lhendrik@science.uva.nl}}

\begin{document}
\unitlength=.25em
\maketitle

\begin{abstract}
The non-classical, nonmonotonic inference relation associated with the
answer set semantics for logic programs gives rise to a relationship of 
{\em strong equivalence} between logical programs that can be verified in 
3-valued G\"{o}del logic, {\bf G3}, the strongest non-classical intermediate 
propositional logic~\cite{lpv}. In this paper we will show that {\bf KC} 
(the logic obtained by adding axiom  $\neg A\vee\neg\neg A$ to intuitionistic 
logic), is the weakest intermediate logic for which strongly equivalent logic 
programs, in a language allowing negations, are logically equivalent.

\end{abstract}

\section{Introduction}
In logic programming certain fragments of first-order logic are given
a computational meaning. The first
and best known example of such a fragment is that of the {\em Horn
clauses}, quantifier-free formulas
of the form $B$ or $\rwedge A_i\imp B$, where the $A_i$ and $B$ are atomic
formulas, the basis for the
programming language Prolog~\cite{kow}. A logic program is a finite set of
such formulas (called {\em rules}).

A logic program $\Pi$ in the language $L$ can be interpreted in first-order
logic as a set of sentences in $L$, by taking the universal closure 
$\forall\vec{x} R$ of each of the rules $R$ in $\Pi$. An
alternative method of eliminating the free variables in the rules, is
the substitution of {\em ground terms}, i.e.\
terms built from the constants and functions in $L$. The set of ground
terms of $L$, the {\em Herbrand
Universe} of $L$, may be used as the domain for models of theories in $L$,
the {\em Herbrand models}.
Replacing each $R$ by the set of all possible substitutions with ground
terms, $\Pi$ is now replaced by
a set of quantifier-free sentences in $\Pi^H$ in $L$.
The rationale behind this is given by the following well-known fact
(for a proof see for example~\cite{doets}).

\begin{fact}
Let $\Pi$ be a set of universal sentences. The following are equivalent:
\begin{enumerate}
\item $\Pi$ has a model
\item $\Pi$ has a Herbrand model
\item $\Pi^H$ is satisfiable in propositional logic
\end{enumerate}
\end{fact}

In this paper we are interested in logic programs as (possibly infinite)
sets of propositional formulas.
In general, our program rules may include negations and disjunctions and
hence the results in this paper extend to disjunctive logic programming as well.

We will denote a fragment of the language of propositional logic by
enumerating between square brackets  the connectives and constants
that are allowed in formulas of the fragment. So $[\wedge,\neg]$ will
denote
the set of formulas built from atomic formulas, using only
conjunction and negation. And $[\wedge,
\vee,\bot,\top]$ will be the fragment of formulas built with
conjunction and disjunction from atomic formulas
and the constants $\bot$ and $\top$.

As long as we remain in classical propositional logic, a model $w$ for a
program $\Pi$ can be identified by the set of atoms $X$ valid in $w$, in our 
notation $w=\langle X\rangle$.

In logic programming one is interested in constructing a most
'general' model for a program $\Pi$ (in the language $L$ based on the Herbrand 
Universe of $L$). Such a 'most general' model should not identify terms, for 
example, unless such an identity is implied by $\Pi$.
For the propositional analogue of programs with Horn clauses as rules,
the {\em minimal Herbrand model} is such a 'most general' model.

In propositional logic an obvious candidate for the most general model
of $\Pi$ is the intersection of
all sets of atoms $X$ such that $\langle X\rangle\models\Pi$ (i.e.,
$\Pi$ is true in the valuation that makes exactly the atoms in $X$ true). 
For programs with Horn clauses as rules this works fine as can be seen from 
the following fact.

\begin{fact}
Let $\Pi\subseteq \{A\imp B\mid A, B\in [\wedge,\bot,\top]\}$ and
$X=\bigcap\{Y\mid \langle Y\rangle\models\Pi\}$. Then $\langle
X\rangle\models\Pi$.
\end{fact}

Although the above fact introduces a fragment slightly richer than the
language of Horn clauses, it is
still easy to prove.

A simple example, like $p\vee q$, shows that there may not be a unique
minimal model for $\Pi$ if disjunctions are allowed in the rules of $\Pi$. 
And even more serious problems arise for the notion of {\em most general model}, 
when negations in the head or body of rules of $\Pi$ are allowed.

Several solutions have been proposed for the semantics of logic programs
with disjunctions and negations. The answer set (or 'stable model') semantics  
we use in this paper was introduced by Gelfond and Lifschitz
in~\cite{g&l}. The main idea\footnote{A more precise definition will be
given in the sequel.} is that $X$ is an {\em answer set} of $\Pi$ if $\langle
X\rangle\models\Pi^X$ where $\Pi^X$ is the program that arises if we
replace all negations $\neg A$ in $\Pi$ by either $\bot$ or $\top$ according to
whether $\langle X\rangle\models A$ or not, and that $X$ is minimal in this 
respect.
The behavior of negation in this semantics resembles that of the {\em
negation as failure} (or {\em negation by default}) in many Prolog implementations.

We define a logic {\bf L} to be {\em sound for stable models} or {\em
sound for stable inference} if, whenever $\Pi\vdash_{\bf L} A$, then
the answer sets for $\Pi\cup\{A\}$ are the same as for $\Pi$.
Classical propositional logic, {\bf CPL}, turns out to be too strong
to have this property. For example, the program $\neg\neg p$ does not
have answer sets at all, whereas $\neg\neg p\vdash_{\bf CPL}
p$ and $\{\neg\neg p, p\}$ has $\{p\}$ as its answer set.

Logics weaker than classical logic {\bf CPL} can give a solution to
this problem, in particular {\em intermediate logics}, i.e.\ logics
derived from intuitionistic propositional logic {\bf IPL} by adding axioms 
that are valid in {\bf CPL}, do provide a sound basis for stable
inference. David Pearce used in~\cite{pearce} 3-valued G\"{o}del
logic {\bf G3} to prove logic programs strongly equivalent.
The notion of {\em strong equivalence} of logic programs was
introduced in \cite{lpv}. Logic programs
$\Pi_1$ and $\Pi_2$ are said to be {\em strongly equivalent} in the
sense of stable model semantics
if for every logic program $\Pi$, $\Pi_1\cup\Pi$ and $\Pi_2\cup\Pi$
have the same answer sets.
For programs in the language $\{A\imp B\mid A, B\in
[\wedge,\vee,\neg]\}$ it was shown in~\cite{lpv}
that $\Pi_1$ and $\Pi_2$ are strongly equivalent precisely if they are
equivalent in {\bf G3}.

As pointed out in \cite{lpv}, the notion of strong equivalence may be
of interest in showing that a part
of a program can be replaced by a simpler equivalent part, without
affecting the behavior of the whole
program or its extensions. Replacing nonclassical, nonmonotonic stable
inference by a well-understood monotonic intermediate logic like {\bf G3}, 
will simplify the verification of such strong equivalence between logic 
programs. The logic {\bf G3}, also known as the Smetanich logic of 
{\em here-and-there}, is the intermediate logic whose models are based on the 
partially ordered frame $\langle h, t\rangle$ with $h\leq t$ (\cite{pearce}, 
\cite {cz}, and see section~\ref{prel}).

In this paper we will consider the problem of the 'weakest'
intermediate logic {\bf L} for which provable equivalence is the same
as strong equivalence in the sense of stable models. 
In other words, which {\bf L} has the property that logic programs 
$\Pi_1\cup\Pi$ and $\Pi_2\cup\Pi$ have the same answer sets for all $\Pi$ iff 
$\Pi_1$ and $\Pi_2$ are equivalent in {\bf L}, but this property does not hold
for any strictly weaker logic. Note that this will depend on the language 
one allows for the programs.
Our main result is that for programs in the language $\{A\imp B\mid A,
B\in [\wedge,\vee,\neg]\}$ the weakest intermediate logic for which equivalence 
of programs equals strong equivalence on stable models is
the logic {\bf KC}, axiomatized by adding axiom $\neg A\vee\neg\neg A$
to {\bf IPL}. This logic (also known as {\em Jankov's logic} or the
{\em logic of the weak law of excluded middle}) was introduced in \cite{ja}.

Our main result remains true if we restrict the language of programs to
$\{A\imp B\mid A, B\in [\vee,\neg]\}$ or $\{A\imp B\mid A, B\in
[\wedge,\neg]\}$.
But in the language $\{A\imp B\mid A, B\in [\wedge,\vee,\bot,\top]\}$
strong equivalence of programs
coincides with equivalence in {\bf IPL} itself.

Let us note that {\bf G3} is easier to implement than {\bf KC}. This
is witnessed by the fact that satisfiability in {\bf G3} is $NP$ and
satisfiability in {\bf KC} is $PSPACE$. However, in many
particular cases it is easy to see that certain formulas are not
derivable in {\bf KC} whereas this is a complex matter for {\bf
G3}. This point also shows up when one wants to prove that the
disjunctive rule $p\vee q$ is not strongly equivalent to any
nondisjunctive rule. It is not clear how this could be done using the
characterization of strong equivalence as provable equivalence in {\bf
G3} since $\vee$ is definable from the other connectives in {\bf
G3}. But with our characterization of strong equivalence as provable
equivalence in {\bf KC} it is a rather simple corollary which we will prove at 
the end of the paper.

\section*{Acknowledgements}
We would like to thank David Pearce who challenged us to find the weakest 
intermediate logic for which equivalent programs are strongly equivalent.
We are also obliged to him and V. Lifschitz for diligently explaining some of 
the subtleties in answer set semantics. Finally the anonymous referees should 
be acknowledged for their corrections and valuable suggestions for improving 
the presentation.

\section{Preliminaries}\label{prel}
In the language of propositional logic formulas are built from atoms
(plus possibly constants $\top$ and $\bot$) using $\wedge,\vee,\imp$, $\neg$. 
Fragments of propositional logic are obtained by restricting
the use of atoms, constants and/or the use of the connectives.

The Kripke semantics in this paper is fairly standard.

\begin{definition} A {\em Kripke frame} $\langle W,
\leq\rangle$ is a set of worlds (or nodes)
$W$ with a partial ordering $\leq$. A {\em model} $M$ will be such
a frame together with a function $\mbox{atom}(w)$ mapping each
world $w\in W$ to a set of atomic formulas, such that if $w\leq v$ then
$\mbox{atom}(w)\subseteq\mbox{atom}(v)$.
\end{definition}

Note that Kripke models are not necessarily rooted. A {\em maximal world} $w\in W$ 
(i.e.\ such that for all $v\in W$, $w\leq v$ implies $v=w$) will be called a 
{\em terminal node} of $W$ (or $M$).

For the language of propositional logic the interpretation in a world
$w$ of a model $M$ (by which we mean $w\in W$ if
$M=\langle W,\leq\rangle$) is given by the usual rules.

\begin{definition}
$w\models_M A$ ($A$ is true in $M$ at $w$) is defined by recursion on
the length of $A$.
\begin{enumerate}
\item $w\models_M p\iffi p\in {atom}(w)$,
\item $w\models_M A\wedge B\iffi w\models_M A$ and $w\models_M B$,
\item $w\models_M A\vee B\iffi w\models_M A$ or $w\models_M B$,
\item $w\models_M A\imp B\iffi\forall v\geq w\, (v\not\models_M A$ or
$v\models_M B)$.
\item $w\models_M\top$,
\item $w\not\models_M\bot$,
\end{enumerate}
\end{definition}

If it is clear from the context which model is meant in $w\models_M
A$, we will omit the subscript (and simply write $w\models A$). 
If $T$ is a set of formulas and $w$ a world in a Kripke model $M$, then
$w\models T$ iff $w\models A$ for all $A\in T$. We will write
$M\models A$ (or $M\models T$) if for all $w$ in $M$ it is true that 
$w\models A$ (or $w\models T$).
A well-known fact about Kripke models is that if $w\models A$ and
$w\leq v$ then $v\models A$, which is true for atomic formulas $A$ by the
monotonicity of the function $\mbox{\em atom}$ but extends to all formulas $A$.

Intuitionistic propositional logic ({\bf IPL}) is sound and complete
for the set of finite Kripke models. Thus, $\vdash_{\bf IPL} A$ iff 
$M\models A$ for each finite $M$.

Classical propositional logic ({\bf CPL}) is sound and complete for
the set of Kripke models where the partial ordering is identity. 
Hence, a classical model consists of a world $w$ that may be
identified with $\mbox{\em atom}(w)$, the set of atomic formulas valid in $w$.
If $\mbox{\em atom}(w)=X$ we will denote $w\models A$ by $\langle
X\rangle\models A$.
In the case where $w$ is a node in a Kripke model $M$, $\langle w\rangle$
will denote the classical world with the same set of atoms as the node $w$ 
(so $\langle w\rangle = \langle\mbox{\em atom}(w)\rangle$).

An {\em intermediate logic} is a logic obtained by adding formulas valid in
{\bf CPL}, to {\bf IPL} as schemes.

3-valued G\"odel logic {\bf G3} can be defined as the logic sound and
complete for models based on the frame
$\langle\{ h, t\},\leq\rangle$ with $h\leq t$ (in a short notation:
$\langle h, t\rangle$). We will call these models {\em here-and-there
models}. {\bf G3} traditionally is introduced by giving the (3-valued)
truth tables for the connectives. The three values correspond in the context
of Kripke models of course to the three sets of nodes that a formula
can be true in: $\emptyset,\{t\},\{h,t\}$. Alternatively {\bf G3} may
be obtained by adding e.g.\ one of the following axioms to {\bf IPL}:
\begin{enumerate}
\item\label{luk} $(\neg A\imp B)\imp(((B\imp A)\imp B)\imp B)$
\item $(A\leftrightarrow B)\vee(A\leftrightarrow C)\vee(A\leftrightarrow
D)\vee(B\leftrightarrow C)
\vee(B\leftrightarrow D)\vee(C\leftrightarrow D)$
\item\label{hos} $A\vee(A\imp B)\vee\neg B$
\item $(((A\imp(((B\imp C)\imp B)\imp B))\imp A)\imp A)\wedge(\neg A
\vee\neg\neg A)$
\end{enumerate}
\L ukasiewicz~\cite{lu} seems to have been the first to axiomatize {\bf
G3}, using axiom~\ref{luk}.
The second axiom is G\"{o}del's~\cite{go} formula expressing that there are
only three truth values.
The third is a simplified version of Hosoi's axiom $A\vee\neg
A\vee(A\imp B)\vee(B\imp C)$~\cite{ho}.
The last axiom is a combination of the iterated Peirce formula (the
substitution of the Peirce formula
$((B\imp C)\imp B)\imp B$ for $B$ in $((A\imp B)\imp A)\imp A$ ) and
the axiom for {\bf KC} (see below),
together expressing that the logic will be complete with respect to
frames of maximal depth 2 and a single
terminal node. Clearly $\neg A\vee\neg\neg A$ can also easily be
derived from~\ref{hos} (take $B=\neg A$
and use that $A\imp\neg A$  and $\neg A$ are equivalent and
$A\vee\neg\neg A$ is equivalent to $\neg\neg A$) or the other axioms.
For more details see~\cite{cz}.

We will use the notation $\langle Y, X\rangle$ for the Kripke model
$\langle h, t\rangle$,
with $X=\mbox{\em atom}(t)$ and $Y=\mbox{\em atom}(h)$.

\bigskip
The intermediate logic {\bf KC} is given by the rules and axioms of
intuitionistic propositional logic {\bf IPL} plus the axiom 
$\neg A\vee \neg\neg A$. {\bf KC} is sound and complete with
respect to the finite (rooted) Kripke models with a single terminal
node (\cite{ja}, see~\cite{cz}).

The Kripke models of {\bf G3} are a special kind of {\bf KC}-Kripke
models, hence by the soundness and completeness theorems for {\bf G3} and 
{\bf KC}, provability (from a set of formulas $T$) in {\bf KC} implies 
provability (from $T$) in {\bf G3}: $T\vdash_{\bf KC}~A$ implies 
$T\vdash_{\bf G3} A$.

\section{Answer sets and stable models}
In this section we recall some of the definitions and results
from~\cite{g&l} and~\cite{ltt} for
programs in the language $\{A\imp B\mid A, B\in
[\wedge,\vee,\neg]\}$. As in~\cite{lpv} our language allows
more complex rules than the usual $A_1\wedge\ldots\wedge A_m\wedge\neg
A_{m+1}\ldots\wedge\neg A_n\imp
B_1\vee\ldots\vee B_k$ (conjunctions, disjunctions and negations can
be nested). 
We will try to state and prove the results for as large a class
of formulas as possible.

We will start with some results for programs in the language $\{A\imp B\mid
A,B\in [\wedge, \vee,
\bot,\top]\}$. Examples of rules in this language are: $p$ (i.e.\ $\top\imp
p$), $p\wedge q\vee r\imp
p\wedge r$ and $p\imp\bot$. Note that also negations of formulas $A$ in
$[\wedge, \vee]$ are allowed, as rules, if $\neg A$ is written as $A\imp\bot$.

\begin{definition}\label{as1}
Let $\Pi$ be a program in $\{A\imp B\mid A,B\in [\wedge, \vee,
\bot,\top]\}$. A set of atoms $X$ is
an {\em answer set} of $\Pi$ if for all $Y\subseteq X$ it is true that
$\langle Y\rangle\models\Pi
\iffi Y=X$.
\end{definition}

A program in $\{A\imp B\mid A,B\in [\wedge, \vee, \bot, \top]\}$ may
have several answer sets
(like for example the program $p\vee q$) and (logically) different programs
may have the same
answer sets (for example $p\imp q$ and $q\imp p$ both have the empty set as
their only answer set).

\begin{definition}
Programs $\Pi_1$ and $\Pi_2$ in $L$ are called {\em strongly equivalent}
(in $L$) if for every program
$\Pi$ in $L$ the programs $\Pi_1\cup\Pi$ and $\Pi_2\cup\Pi$ have the
same answer sets.
\end{definition}

Logic programs in $\{A\imp B\mid A,B\in [\wedge, \vee, \bot, \top]\}$ are
strongly equivalent if and
only if (viewed as sets of propositional formulas) they are equivalent
in classical propositional logic.

\begin{theorem}\label{t1}
Let $L=\{A\imp B\mid A,B\in [\wedge, \vee, \bot,\top]\}$ and let $\Pi_1$
and $\Pi_2$ be programs in $L$.
$\Pi_1$ and $\Pi_2$ are strongly equivalent if they are equivalent in
{\bf CPL}, i.e.\ $\Pi_1\equiv_{\bf CPL} \Pi_2$.
\end{theorem}

\begin{proof} First assume $\Pi_1\equiv_{\bf CPL} \Pi_2$ and let $X$ be
an answer set for $\Pi_1\cup\Pi$.
Then for $Y\subseteq X$ with $\langle Y\rangle\models \Pi_2\cup\Pi$ we
may infer that
$\langle Y\rangle\models \Pi_1\cup\Pi$ and hence $Y=X$. Which proves $X$ is
also an answer set for
$\Pi_2\cup\Pi$. Likewise, every answer set for $\Pi_2\cup\Pi$ can be proven
to be an answer set for
$\Pi_1\cup\Pi$ and hence $\Pi_1$ and $\Pi_2$ are strongly equivalent.

For the other direction, let $\langle X\rangle\models\Pi_1$ and let
$\Pi=X$. Observe that $X$ is an
answer set for $\Pi_1\cup\Pi$ and, as $\Pi_2$ is strongly equivalent to
$\Pi_1$, $X$ is also an
answer set for $\Pi_2\cup\Pi$. Which proves $\langle X\rangle\models\Pi_2$.
Likewise, every model of
$\Pi_2$ will be a model of $\Pi_1$, which proves $\Pi_1\equiv_{\bf
CPL}\Pi_2$.
\end{proof}

For a more general treatment of negations in logic programs the
following {\em reduction} of a program was introduced in~\cite{g&l},\cite{ltt}.

\begin{definition}\label{ax}
Let $X$ be a set of atomic formulas and $A$ a formula. $A^X$ is defined
recursively as:

\bigskip
$\begin{array}{lll}
p^X & = p & \mbox{if}\; p\;\mbox{is atomic}\\
(A\circ B)^X & = A^X\circ B^X & \mbox{for}\;\circ\in\{\wedge,\vee,\imp\}\\
(\neg A)^X & = \left\{\begin{array}{ll}
\bot & \mbox{if}\; \langle X\rangle\models A\\
\top & \mbox{otherwise}\\
\end{array}\right.
\end{array}$
\end{definition}

For a program $\Pi\subseteq\{A\imp B\mid A, B\in [\wedge,\vee,\neg]\}$ 
the reduction $\Pi^X$ will be a program in 
$\{A\imp B\mid A,B\in [\wedge, \vee,\bot,\top]\}$.

\begin{definition}\label{as2}
Let $\Pi\subseteq \{A\imp B\mid A, B\in [\wedge,\vee,\neg]\}$. A set $X$ of
atomic formulas
is called an {\em answer set} for $\Pi$ if for all
$Y\subseteq X$ we have $\langle Y\rangle\models\Pi^X\iffi Y=X$.
\end{definition}

If we restrict the language to $\{A\imp B\mid A, B\in
[\wedge,\vee,\bot,\top]\}$, we have
$\Pi^X=\Pi$ and definition~\ref{as2} coincides with definition~\ref{as1}.

To find a theorem similar to theorem~\ref{t1} for strong equivalence in
$\{A\imp B\mid A,B\in
[\wedge, \vee, \neg]\}$, we will use the characterization of answer sets
in~\cite{pearce}, based
on Kripke models for the intermediate logic {\bf G3}.

The following lemma is not only useful in this case but also will have
applications in the next
section. Recall that for a world $w$ in a Kripke model $M$, $\langle
w\rangle$ denotes the
classical model $\langle\mbox{\em atom}(w)\rangle$.

\begin{lemma}\label{wv}
Let $w$ be a node in a Kripke model $M$.
For $A, B\in [\wedge,\vee,\bot,\top]$, $w\models A$ iff $\langle
w\rangle\models A$ and,
if $w\models A\imp B$ then $\langle w\rangle\models A\imp B$.
\end{lemma}

\begin{proof}
First we prove for $A\in [\wedge,\vee,\bot,\top]$ that
$w\models A$ iff
$\langle w\rangle\models A$. If $A$ is atomic, $\bot$ or $\top$, this is
obvious. By induction
on the complexity of $A$, the proof for the cases of conjunction and
disjunction is straightforward.

For the second part of the proof, let both $A$ and $B$ be in
$[\wedge,\vee,\bot,\top]$.
If $\langle w\rangle\not\models A\imp B$ then $\langle w\rangle\models A$
and $\langle w\rangle
\not\models B$. By the first part of the lemma then $w\not\models A\imp
B$.
\end{proof}

As an immediate consequence of lemma~\ref{wv} we have the following lemma
for models of {\bf G3}.

\begin{lemma}\label{01}
For $A, B\in [\wedge,\vee,\bot,\top]$, $\langle Y, X\rangle\models A\imp B$ iff
$\langle X\rangle\models A\imp B$ and $\langle Y\rangle\models A\imp B$
\end{lemma}

\begin{proof} 
Assume $\langle Y, X\rangle\models A\imp B$. By lemma~\ref{wv}
we may conclude that
$\langle X\rangle\models A\imp B$ and $\langle Y\rangle\models A\imp B$.

For the other direction, assume $\langle X\rangle\models A\imp B$ and
$\langle Y\rangle\models
A\imp B$. If $\langle Y, X\rangle\models A$, then $\langle Y\rangle\models
A$ and hence
$\langle Y\rangle\models B$, which implies $\langle Y, X\rangle\models B$, so
$\langle Y, X\rangle\models A\imp B$. On the other hand if $\langle Y,
X\rangle\not\models A$
then $\langle X\rangle\models A\imp B$ immediately implies $\langle Y,
X\rangle\models A\imp B$.
\end{proof}

The next lemma is true for all propositional formulas.

\begin{lemma}\label{x3}
For all sets of atoms $X$ and $Y$ such that $Y\subseteq X$ it is true that
$\langle Y, X\rangle\models A \iffi\langle Y, X\rangle\models A^X$.
\end{lemma}

\begin{proof}
Observe that $\langle Y,X\rangle\models\neg A \iffi \langle
X\rangle\not\models A$.
As a consequence we have  $\langle Y,X\rangle\models\neg A \iffi (\neg A)^X
=\top \iffi
\langle Y,X\rangle\models (\neg A)^X$.
Hence for all $A$ it is true that $\langle Y,X\rangle\models \neg A\equi
(\neg A)^X$.
This implies, using the definition~\ref{ax}, that for all
$A\in[\wedge,\vee,\imp,\neg]$ it is true that $\langle Y,X\rangle\models
A\equi A^X$, from which
the lemma immediately follows.
\end{proof}

Theorem~\ref{p1}, theorem~\ref{sim} and corollary~\ref{lpv} restate the
main result of~\cite{lpv}.

\begin{theorem}\label{p1}
Let $\Pi\subseteq \{A\imp B\mid A, B\in [\wedge,\vee,\neg]\}$ and $X$ a set
of atomic formulas.
$X$ is an an answer set of $\Pi$ if and only if for all $Y\subseteq X$
it is true that
$\langle Y,X\rangle\models\Pi\iffi X=Y$.
\end{theorem}

\begin{theorem}\label{sim}
Let $\Pi_1$ and $\Pi_2$ be programs in $\{A\imp B\mid A, B\in
[\wedge,\vee,\neg]\}$.
$\Pi_1$ and $\Pi_2$ are strongly equivalent if and only if they are
equivalent in {\bf G3},
i.e.\ $\Pi_1\equiv_{\bf G3}\Pi_2$.
\end{theorem}

\begin{coroll}\label{lpv}
Let $\Pi_1$ and $\Pi_2$ be programs in $\{A\imp B\mid A, B\in
[\wedge,\vee,\neg]\}$.
$\Pi_1$ and $\Pi_2$ are strongly equivalent if and only if for all
$\Pi\subseteq
\{p\imp q\mid p \;\mbox{atomic or}\; p=\top, q \;\mbox{atomic}\}$,
$\Pi_1\cup\Pi$ and $\Pi_2\cup\Pi$ have the same answer sets.
\end{coroll}

According to the corollary above, the notion of strong equivalence of
logic programs may depend
on the language for the programs $\Pi_1$ and $\Pi_2$, but in all
sublanguages $L$ of
$\{A\imp B\mid A, B\in [\wedge,\vee,\neg]\}$, we may use theorem~\ref{sim},
as long as
rules of the form $p\imp q$ (with $p$ and $q$ atomic) are in $L$.

\section{Stable inference in intermediate logics}
The previous section linked strong equivalence of logic programs in stable
inference with
equivalence in {\bf CPL} (for programs without negations in the head or the
body of the
rules) or in {\bf G3}. In this section we will determine for several
fragments of propositional logic
 the weakest intermediate logic for which equivalence of programs is
implied by strong
equivalence in stable inference.

For the fragment $\{A\imp B\mid A, B\in [\wedge,\vee,\bot,\top]\}$ we have
the following
lemma.

\begin{lemma}\label{wv2}
Let $L=\{A\imp B\mid A, B\in [\wedge,\vee,\bot,\top]\}$ and $T\subseteq L$,
$C\in L$.
Then $T\vdash_{\bf CPL} C\iffi T\vdash_{\bf IPL} C$.
\end{lemma}

\begin{proof}
The direction from {\bf IPL} to {\bf CPL} is trivial. So let us assume
$T\vdash_{\bf CPL} C$. Let $M$ be a Kripke model and $w$ a node in $M$ such
that $w\models T$. Using lemma~\ref{wv} we may conclude $\langle w\rangle\models
T$ and hence $\langle w\rangle\models C$. Again, use lemma~\ref{wv} to prove 
$w\models C$. This proves that $T$ implies $C$ in Kripke models in general and 
hence $T\vdash_{\bf IPL} C$.
\end{proof}

\begin{coroll}
Let $L=\{A\imp B\mid A, B\in [\wedge,\vee,\bot,\top]\}$ and $\Pi_1,\Pi_2
\subseteq L$.
Then $\Pi_1$ and $\Pi_2$ are strongly equivalent in $L$ iff
$\Pi_1\equiv_{\bf IPL}\Pi_2$.
\end{coroll}

\begin{proof}
Is an immediate consequence of theorem~\ref{t1} and lemma~\ref{wv2}.
\end{proof}

Strong equivalence between programs in the language $\{A\imp B\mid A,
B\in [\wedge,\vee,\neg]\}$ will not be the same
as equivalence in {\bf IPL}, as for example $\neg p\vee\neg\neg p$ is
strongly equivalent to $\top$
(it is a derivable formula in {\bf G3}) and is not derivable in {\bf IPL}.
The intermediate logic {\bf KC}, which has $\neg A\vee\neg\neg A$ as its axiom, 
will, in the following, turn out to be the weakest intermediate
logic for which equivalence of programs is implied by strong
equivalence in answer set semantics.

\begin{lemma}
Let $L=\{A\imp B\mid A, B\in [\wedge,\vee,\neg]\}$, $T\subseteq L$ and
$C\in L$. Then
$T\vdash_{\bf KC} C\iffi T\vdash_{\bf G3} C$.
\end{lemma}

\begin{proof}
Again the direction from {\bf KC} to {\bf G3} is trivial. For the other 
direction, let $T\nvdash_{\bf KC} A\imp B$ (where $A, B\in[\wedge,\vee,\neg]$). 
Then for some Kripke model $M$ with a single terminal (i.e.\ maximal) node $t$, 
there is a $w\in M$ such that $w\models T, w\models A$ and
$w\not\models B$. We will prove that for the {\bf G3}-model $\langle w, t\rangle$ 
we have for all formulas $C\in[\wedge,\vee,\neg]$ that 
$w\models C\iffi \langle w,t\rangle\models C$ and for $C, D\in[\wedge,\vee,\neg]$ 
that $w\models C\imp D\Imp \langle w,t\rangle\models C\imp D$. As a
consequence, $\langle w,t\rangle\models T, \langle w, t\rangle\models A$ and
$\langle w,t\rangle\not\models B$, which proves $T\nvdash_{\bf G3} A\imp B$.

The proof that for $C\in[\wedge,\vee,\neg]$ we have $w\models C\iffi
\langle w,t\rangle\models C$ is by structural induction. For atomic formulas it 
is obvious and the cases for conjunctions and disjunctions
are trivial. 
For the case of negation, observe that $w\models\neg C\iffi t\not\models C$, and
$t\not\models C\iffi \langle t\rangle\not\models C\iffi \langle
w,t\rangle\models\neg C$.

Now let $C, D\in[\wedge,\vee,\neg]$ and $w\models C\imp D$. Since
$w\leq t$, $\langle t\rangle
\models C\imp D$. So, if $w\not\models C$, we have (by the above part
of the proof) $\langle w, t\rangle
\not\models C$ and hence $\langle w,t\rangle\models C\imp D$. On the
other hand, if $w\models C$, then
also $w\models D$ and by the above part of the proof, also $\langle
w,t\rangle\models C\imp D$.
Which proves that $w\models C\imp D$ implies $\langle
w,t\rangle\models C\imp D$.
\end{proof}

\begin{coroll}\label{colkc}
Let $L=\{A\imp B\mid A, B\in [\wedge,\vee,\neg]\}$ and $\Pi_1,\Pi_2
\subseteq L$.
Then $\Pi_1$ and $\Pi_2$ are strongly equivalent in $L$ iff
$\Pi_1\equiv_{\bf KC}\Pi_2$.
\end{coroll}

\begin{coroll}
{\bf KC} is the weakest intermediate logic {\bf L} such that
$\Pi_1,\Pi_2\subseteq\{A\imp B\mid A, B\in [\wedge,\vee,\neg]\}$ are 
strongly equivalent iff $\Pi_1\equiv_{\bf L}\Pi_2$.
\end{coroll}

\begin{proof}
Note that the {\bf KC} axiom $\neg A\vee\neg\neg A$ can be
expressed in the language
$\{A\imp B\mid A, B\in [\wedge,\vee,\neg]\}$.
\end{proof}

That {\bf KC} is the weakest intermediate logic for strong equivalence
in almost any language with
negation (where negation is taken to be a {\em negation by default}
and strong equivalence defined
according to the answer set semantics) can be seen from the following
corollary.

\begin{coroll}
Let $L=\{A\imp B\mid A,B\in[\neg]\}$ and $\Pi_1,\Pi_2 \subseteq L$.
Then $\Pi_1$ and $\Pi_2$ are strongly equivalent in $L$ iff
$\Pi_1\equiv_{\bf KC}\Pi_2$.
\end{coroll}

\begin{proof}
{\bf KC} can alternatively be axiomatized as {\bf IPL}
plus $((\neg A\imp B)\wedge (\neg\neg
A\imp B)\imp B$. In one direction this is clear from the fact that
this axiom immediately follows from $\neg A\vee\neg\neg A$, for the
other direction substitute $\neg A\vee\neg\neg A$ for $B$ in the axiom,
and $\neg A\vee\neg\neg A$ follows. So, the programs $\{q\}$ and
$\{\neg p\imp q,\neg\neg p\imp q\}$ are strongly
equivalent and any logic making such programs equivalent will be as
strong as {\bf KC}.
\end{proof}

As a consequence also strong equivalence in for example $\{A\imp B\mid A,
B\in [\wedge,\neg]\}$ and
$\{A\imp B\mid A, B\in [\vee, \neg]\}$ will coincide with equivalence in
{\bf KC}.

Even if we restrict the language further, allowing in the body only atoms
or negated atoms and in the
head only atoms (apart from simple statements of atoms and negation of
atoms), {\bf KC} is still the
weakest intermediate logic {\bf L} such that equivalence of programs
in {\bf L} corresponds with
strong equivalence.

In logic programming the programs in this restricted language are known as
{\em normal} programs and have historically been most important.
Most Prolog implementations of {\em negation by default} are restricted to 
this kind of programs, often called {\em general programs} in this context 
(see~\cite{doets}).   

\begin{definition}
A {\em normal} logic program is a finite set of rules $\rwedge l_i\imp p$, 
where the $l_i$ are literals (so either atomic or a negation of an atomic
formula) and $p$ is atomic.
\end{definition}

First we will prove that an alternative axiomatization of {\bf KC}, in
the language of normal
programs, is possible.

\begin{lemma}\label{qlem}
$A\wedge C\imp D, \neg A\imp B, \neg C\imp B\vdash_{\bf KC} \neg D\imp B$.
\end{lemma}

\begin{proof}
Of course, this can be automatically checked in a tableau
system as in~\cite{afm}, but let us do it from scratch. 
By {\bf KC} we have $\neg A$ or $\neg\neg A$. If $\neg A$, $B$ and hence 
$\neg D\imp B$, is immediate from $\neg A\imp B$. 
So, we can assume $\neg\neg A$. Similarly, we can assume $\neg\neg C$.
By {\bf IPL}, $\neg\neg(A\wedge C)$ follows. Again by {\bf IPL}, 
$A\wedge C\imp D$ now implies $\neg\neg D$, from which again 
$\neg D\imp B$.
\end{proof}

Let $L$ be the set of formulas coding normal logic programs.
As derivability in {\bf KC} implies derivability in {\bf G3}, we can
use lemma~\ref{qlem} to prove
that $\Pi_1=\{p\wedge r\imp s, \neg p\imp q, \neg r\imp q\}$ and
$\Pi_2=\{p\wedge r\imp s, \neg p\imp q, \neg r\imp q, \neg s\imp q\}$ are 
strongly equivalent programs in $L$.

On the other hand it is easily seen that for each intermediate logic {\bf L}
that proves $\Pi_1$ and $\Pi_2$ equivalent, we have 
$A\wedge C\imp D, \neg A\imp B, \neg C\imp B\vdash_{\bf L} \neg D\imp B$.
The following lemma shows that such an {\bf L} has to contain {\bf KC}.

\begin{lemma}\label{normkc}
If {\bf L} is the intermediate logic with, apart from the axioms of 
{\bf IPL}, the axiom
$(A\wedge C\imp D)\wedge(\neg A\imp B)\wedge(\neg C\imp B)\imp 
(\neg D\imp B)$, then {\bf L} is equivalent with {\bf KC}.
\end{lemma}

\begin{proof}
That $\vdash_{\bf L} A$ implies $\vdash_{\bf KC} A$ is a simple consequence 
of lemma~\ref{qlem}. 
For the other direction, let $A:= p$, $B:= \neg p\vee\neg\neg p$,
$C:=\neg p$ and $D:=\bot$ 
in $(A\wedge C\imp D)\wedge(\neg A\imp B)\wedge(\neg C\imp B)\imp 
(\neg D\imp B)$. All the antecedents as well as $\neg D$ are
then derivable, so $\neg p\vee\neg\neg p$ follows.
\end{proof}

The result of the above discussion is summarized in the next corollary.

\begin{coroll}
Let $\Pi_1$ and $\Pi_2$ be normal logical programs. Then $\Pi_1$ and
$\Pi_2$ are strongly equivalent
iff $\Pi_1\equiv_{\bf KC}\Pi_2$. Moreover, {\bf KC} is the weakest
intermediate logic for which provable
equivalence in the logic and strong equivalence of normal logic
programs coincide.
\end{coroll}

\begin{proof}
The first part immediately follows by corollary~\ref{colkc}.
Lemma~\ref{normkc} implies that {\bf KC} is the weakest intermediate
logic for which equivalent normal
programs are strongly equivalent.
\end{proof}

\begin{coroll}
No program in the language $\{A\imp B\mid A,B\in [\wedge,\neg]\}$ is
strongly equivalent to the program $\{p\vee q\}$.
\end{coroll}

\begin{proof}
Recall that {\bf KC} is sound and complete with respect to the finite Kripke 
models with a single terminal node.  

Let the model $M$ be as pictured below, where 
$\mbox{\em atom}(t) = \{p, q \}$, $\mbox{\em atom}(u) =\{p\}$,
$\mbox{\em atom}(v) =\{q\}$ and $\mbox{\em atom}(w) = \emptyset$.

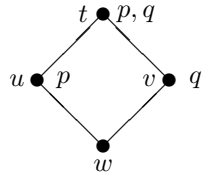
\begin{picture}(25, 30)(-55, 0)
\multiput(15,5)(0,20){2}{\circle*{2}}
\multiput(15,5)(10,10){2}{\line(-1,1){10}}
\multiput(5,15)(20,0){2}{\circle*{2}}
\multiput(15,5)(-10,10){2}{\line(1,1){10}}

\put(12,25){\makebox(0,0){$t$}}
\put(20,25){\makebox(0,0){$p, q$}}

\put(2,15){\makebox(0,0){$u$}}
\put(9,15){\makebox(0,0){$p$}}

\put(22,15){\makebox(0,0){$v$}}
\put(29,15){\makebox(0,0){$q$}}

\put(15,2){\makebox(0,0){$w$}}
\end{picture}

Clearly $u\models p\vee q$ and $v\models p\vee q$, but $w\not\models p\vee q$.
By induction on the complexity of formulas $A\in [\wedge,\imp,\neg]$ one  
easily proves that $$w\models A \iffi u\models A\;\mbox{and}\; v\models A$$

Hence if $\Pi\subseteq \{A\imp B\mid A,B\in [\wedge,\neg]\}$ and $\Pi\equiv_{\bf KC}
\{p\vee q\}$, we would have $u\models\Pi$, $v\models\Pi$, which would imply
$w\models\Pi$, a contradiction.
\end{proof}

Observe that the type of model we need for the proof above is not a {\bf G3}
model (not of the form $\langle h,t\rangle$). In fact, in the full language of
{\bf G3} we can define disjunction using $p\vee q = ((p\imp q)\imp q)\wedge
((q\imp p)\imp p)$. 
The simple proof that this is not possible (in {\bf G3}) if one restricts the 
language of the programs to $\{A\imp B\mid A,B\in [\wedge,\neg]\}$ indicates
that the proof of certain properties of answer set programs may benefit from a
detour in the logic {\bf KC}.

\end{document}